\documentclass[twocolumn]{openjournal}
\usepackage{epsfig}
\usepackage{amsmath}
\usepackage{hyperref}
\hypersetup{
 colorlinks,
citecolor = [rgb]{0,0.5,0} }
\usepackage{amsmath}
\def\gtwid{\mathrel{\raise.3ex\hbox{$>$\kern-.75em\lower1ex\hbox{$\sim$}}}}
\def\ltwid{\mathrel{\raise.3ex\hbox{$<$\kern-.75em\lower1ex\hbox{$\sim$}}}}
\def\\{\hfil\break}

\def\eg{{\it e.g.}}

\def\sun{\hbox{$\odot$}}
\def\lesssim{\mathrel{\hbox{\rlap{\hbox{\lower2pt\hbox{$\sim$}}}\raise2pt\hbox{$<$}}}}
\def\gtrsim{\mathrel{\hbox{\rlap{\hbox{\lower2pt\hbox{$\sim$}}}\raise2pt\hbox{$>$}}}}

\def\arcdeg{\hbox{$^\circ$}}

\newcommand{\unit}[1]{\ifmmode \:\mbox{\rm #1}\else \mbox{#1}\fi}
\begin{document}

\title{Consistency of DES and DESI distances and the Standard Cosmological Model}

\author{Richard Watkins, Cordelia Trueax}
\email{rwatkins@willamette.edu}
\email{cctrueax@willamette.edu}
\affiliation{Department of Physics, Willamette University, Salem, OR 97301, USA}

\author{Hume A. Feldman}
\email{feldman@ku.edu}
\affiliation{Department of Physics \& Astronomy, University of Kansas, Lawrence, KS 66045, USA}

\begin{abstract}
We test the consistency of the cosmic distance-redshift relation inferred from the cosmic microwave background (CMB), the Dark Energy Spectroscopic Instrument (DESI) baryon acoustic oscillation measurements, and the Dark Energy Survey (DES) Type Ia supernovae within the framework of flat $\Lambda$CDM. DESI recovers the CMB-constrained parameter combination $(r_d h)(\Omega_m/0.3)^{0.4}$ with sub-percent precision, demonstrating excellent agreement between BAO measurements at $z \sim 1$ and the acoustic scale at recombination. Imposing the CMB constraint yields an estimate of $\Omega_m$ that is slightly lower than, but only in mild tension with, the Planck value. The high-redshift DES supernova sample is well described by the standard cosmological model, whereas the current low-redshift anchor sample exhibits a systematic offset of $\sim 0.05$ mag that drives much of the apparent preference for evolving dark energy. Preliminary data from the Dark Energy Bedrock All-Sky Supernova Program (DEBASS) do not show this offset, suggesting that unresolved low-redshift systematics may account for the discrepancy. These results suggest that a single flat $\Lambda$CDM model accurately describes the distance-redshift relation from the local Universe to recombination, placing increasingly stringent constraints on new-physics explanations of the Hubble tension.
\end{abstract}

\
\section{Introduction}
\label{sec:intro}

The distance-redshift relation \citep[\eg,][]{Hub1929,Kai88,CanCruCso26}  is one of the most powerful observational probes in cosmology. Measurements of cosmological distances over a wide range of redshifts allow the expansion history of the Universe to be reconstructed and provide direct constraints on the matter content, geometry, and dynamical evolution of the cosmos \citep[\eg,][]{SprMagCol14,SaiColMag20,RieCas2021,HowlettSDSS,CF4-full,ScoRieMur25}. Within the framework of General Relativity \citep{Einstein1916,Friedman1922,Lemaitre1927}, the evolution of the expansion rate is determined by the energy-density components of the Universe \citep[\eg,][]{PPC,Mukhanov2005,ryden2017}, making precise distance measurements a cornerstone of modern cosmological parameter estimation.

Over the past two decades, two observational techniques have played particularly important roles in mapping the distance-redshift relation. Type Ia supernovae \cite[SNIa, see \eg][]{Phillips1993,FreMadHat19,RieYuaMac22,ScoRieMur25,SheAceBro26}) serve as standard candles that trace luminosity distances across cosmic time and provided the first evidence for the accelerated expansion of the Universe \citep{RieFilChal1998,PerAldGol99}. More recently, baryon acoustic oscillations \citep[BAO,][]{EisZehHog05,Cole05,PerColEis07,DESI2025,Efs25b}) have emerged as a complementary standard ruler whose physical scale is calibrated by conditions in the early Universe. Measurements of the BAO scale provide robust determinations of cosmological distances and expansion rates that are largely independent of the astrophysical complexities associated with supernova observations \citep[\eg,][]{MooValRas23,FerMcDBal26}.

The standard spatially flat $\Lambda$CDM cosmological model has been remarkably successful in describing both early- and late-time observations. Measurements of the cosmic microwave background (CMB), particularly those from the Planck satellite \citep{PlanckXLVIII16}, determine cosmological parameters with high precision and predict a specific form for the distance-redshift relation extending from recombination to the present epoch. At lower redshifts, large-scale structure surveys and supernova observations generally support this picture. Nevertheless, increasing observational precision has revealed several areas of tension \citep[\eg,][]{DaiDeSSch2021,HuWang2023,BouPer2024}. Most notably, measurements of the Hubble constant inferred from local distance-ladder techniques differ significantly from values obtained through CMB analyses assuming $\Lambda$CDM \citep[for a review see][]{Di_Valentino2021}. More recently, analyses combining BAO and supernova data have motivated investigations of extensions to the standard model, including evolving dark-energy scenarios. For an extensive review of tensions in cosmology see \citet{CosmoVerse25}.

The recently released measurements from the Dark Energy Spectroscopic Instrument (DESI) provide the most precise BAO constraints to date over a broad redshift range \citep{MooValRas23,DESI2025,Loubser2025}. At the same time, the Dark Energy Survey (DES) has produced one of the largest and most homogeneous samples of Type Ia supernovae, extending luminosity-distance measurements to redshifts approaching unity \citep{AbbAlaAll2019,PopShaKen2026}. Together, these surveys offer an unprecedented opportunity to test the internal consistency of the cosmological distance scale and to assess the extent to which independent probes support the standard cosmological model.

In this paper we examine the consistency of DESI BAO measurements \citep{DESI2025} with the constraints implied by the CMB and then compare the resulting cosmological model with the DES supernova distance measurements \citep{VinKesSha25,GiaAladas2025}. Rather than focusing primarily on parameter estimation, our goal is to determine whether a single $\Lambda$CDM model can simultaneously describe the distance measurements obtained from these independent observational programs. We pay particular attention to the role of the low-redshift supernova sample, which has recently been the subject of discussion regarding possible systematic offsets, and we investigate preliminary results from the Dark Energy Bedrock All-Sky Supernova Program \citep[DEBASS,][]{AceSheBro2026,SheAveBro2026} as an alternative low-redshift anchor.

Our analysis proceeds in three stages. First, in Sec. \ref{sec:DESIcmb}, we test the consistency of DESI BAO measurements with the highly constrained parameter combination determined by the CMB. Second, in Sec. \ref{sec:DESIDES}, we compare the resulting best-fitting cosmological model with the DES supernova data, examining residuals and the influence of the low-redshift sample. Third, in Sec. \ref{sec:DRR}, we investigate the implications of these results for the cosmological distance-redshift relation over the full range from local supernovae to recombination and discuss their relevance to ongoing debates concerning dark energy and the Hubble tension. We present a final discussion and our conclusions in Sec. \ref{sec:DnC}.

\section{Consistency of DESI and Planck}
\label{sec:DESIcmb}

\begin{figure}
\centering
\includegraphics[scale=0.5]{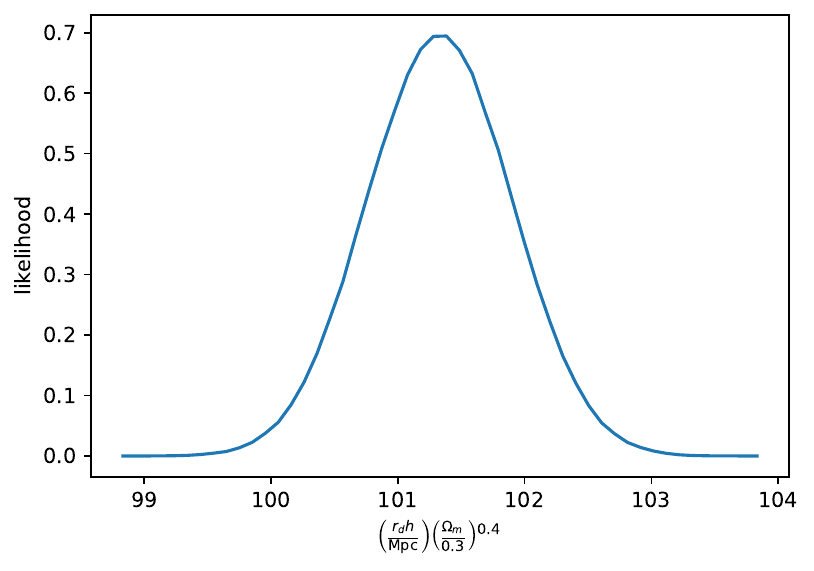}
\caption{The likelihood for the combination of parameters $\left(\frac{r_d h}{\text{Mpc}}\right)\left(\frac{\Omega_m}{0.3}\right)^{0.4}$  calculated using the DESI BAO data, marginalizing over the other independent combination of parameters.   }
\label{fig:combo}
\end{figure}

DESI uses measurements of Baryon Acoustic Oscillations \citep[BAO,][]{DESI2025} to measure $D_M/r_d$, the transverse distance, and $D_H/r_d$, the Hubble distance divided by the drag horizon, at several different redshifts.   Assuming flat $\Lambda$CDM, these measurements can be modeled using the two parameters $\Omega_m$ and the combination $H_or_d$.  However, in the context of this model, Planck puts especially stringent constraints on the combination of both parameters $H_o r_d /\Omega_m^{0.4}$, a constraint that comes from treating the first acoustic peak as a BAO at a redshift of $z\approx 1090$. In particular, Planck finds
\begin{equation}
\left(\frac{r_d h}{\text{Mpc}}\right)\left(\frac{\Omega_m}{0.3}\right)^{0.4} = 101.056\pm 0.036
\end{equation}
This constraint has very small uncertainty and makes almost no assumptions and so makes an excellent first test of consistency.  We can use the DESI BAO data to calculate the likelihood for this combination of parameters using a Markov Chain Monte Carlo (MCMC), marginalizing over the other independent combination.  Given the small uncertainty in the CMB measurement, if the DESI result differed from this value we would have an unambiguous rejection of the standard cosmological model.
The result is given in Fig.~\ref{fig:combo}.   We see that the DESI BAO determine this combination to less than a percent,  with 
\begin{equation}
\left(\frac{r_d h}{\text{Mpc}}\right)\left(\frac{\Omega_m}{0.3}\right)^{0.4} = 101.28\pm 0.57.
\label{eqn:con}
\end{equation}
Thus DESI agrees well with the CMB in regard to this parameter combination.  Another way of looking at this is that a model that goes through the BAO point at $z=1090$ given by the CMB is  also a good fit to the DESI BAO data at redshifts $z\sim 1$.  Given the small uncertainties that DESI places on this parameter combination, we should step back for a moment and consider what a remarkable confirmation of the standard cosmology this is.  

\begin{figure}
\centering
\includegraphics[scale=0.5]{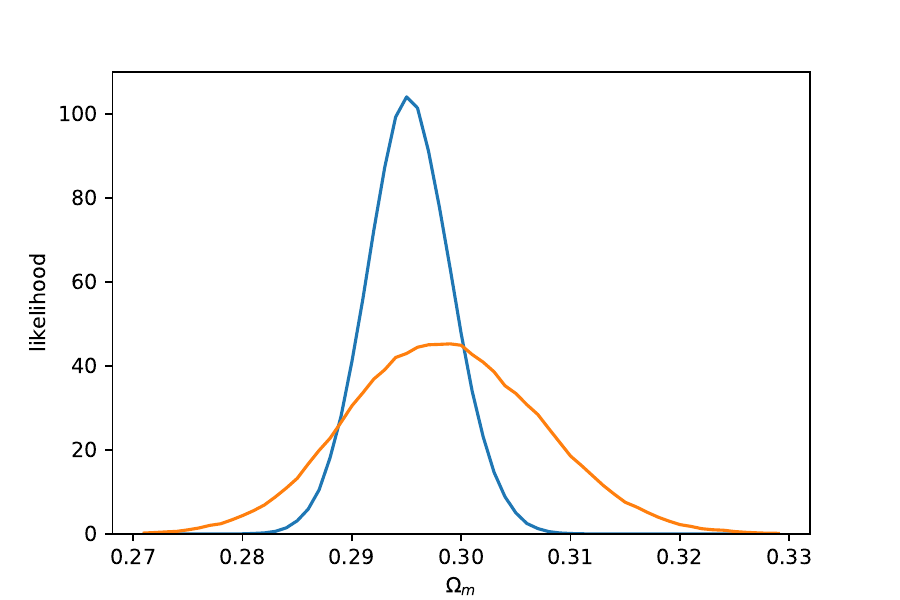}
\caption{The likelihood for $\Omega_m$ (orange curve) marginalizing over other parameters together with the likelihood calculated using the DESI BAO data together with the Planck constraint on the parameter combination $\left(\frac{r_d h}{\text{Mpc}}\right)\left(\frac{\Omega_m}{0.3}\right)^{0.4}$ (blue curve).   }
\label{fig:Om}
\end{figure}

Given the agreement between DESI and the CMB in regard to this parameter combination, it makes sense to further test the standard model by imposing the CMB measurement as a parameter constraint and seeing what this implies for the matter density $\Omega_m$.  Essentially, this keeps us from considering regions of the parameter space that are not possible in the standard cosmological model.  With this constraint, our modeled BAO values depend on only one parameter, $\Omega_m$.  The results of a likelihood analysis are shown in Fig.~\ref{fig:Om}, where we show the $\Omega_m$ likelihood both with and without (marginalizing over the other parameter) the constraint.  We can see that the constraint reduces the uncertainty in $\Omega_m$ significantly but also lowers the central value.  With the constraint, the BAO data yields a value of $\Omega_m= 0.2948\pm 0.0039$.  This is somewhat tighter and lower than the Planck constraint $\Omega_m=  0.3153 \pm 0.0073$; these two values are in tension at about the 2.5 sigma level.  However, recent results from combining ACT with Planck also suggest a lower value of $\Omega_m$ than favored by either ACT or Planck \citep{LouLaPAtk2025}.
It is important to note that since desiwe are using the constraint given in Eqn.~\ref{eqn:con}, $\Omega_m$ and $h r_d$ are not independent parameters.  Indeed, one could equally well view the tension between Planck and DESI as being in the value of $h r_d$.  

\section{Consistency of DESI and DES}
\label{sec:DESIDES}

\begin{figure}
\centering
\includegraphics[scale=0.5]{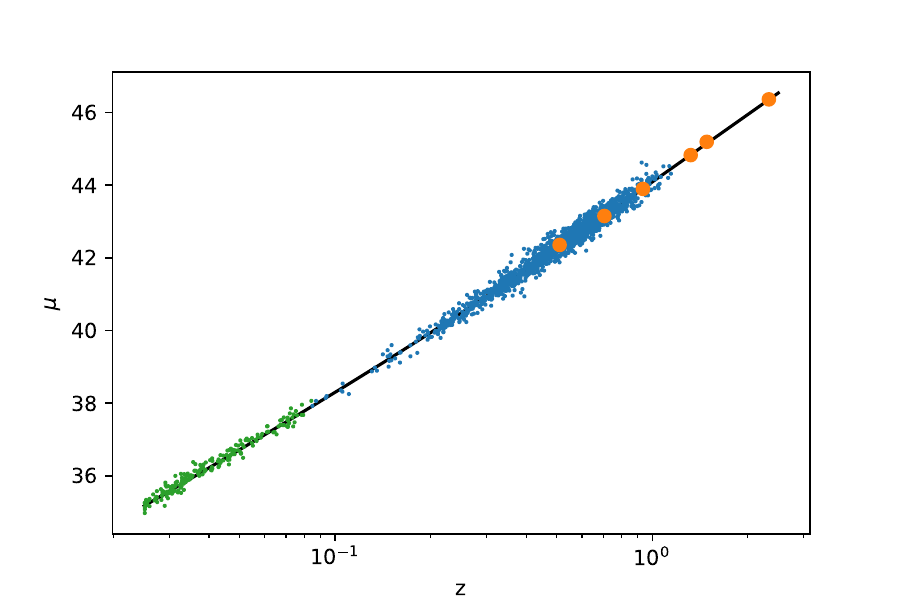}
\caption{Distance modulus $\mu$ vs. redshift for DES supernovae together with the model using the best fit $\Omega_m$ from the DESI BAO and the best fit $H_o$ from the DES sample.  The model is the black line, the low-redshift DES supernovae are shown in green, the high-redshift DES supernovae are shown in blue, and we have included DESI BAO distance measurements (converted to luminosity distances) in orange to show the overlap in the two surveys.  }
\label{fig:distz}
\end{figure}

\begin{figure}
\centering
\includegraphics[scale=0.5]{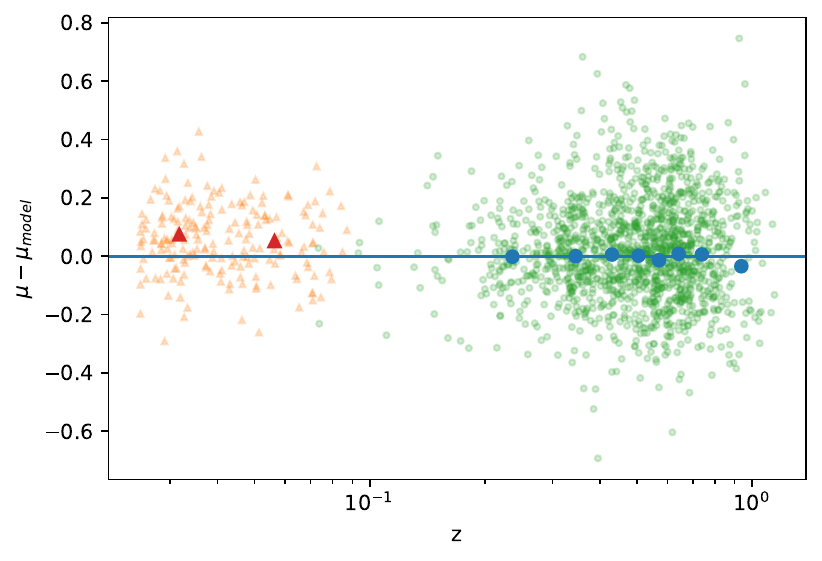}
\caption{Residuals $\mu -\mu_{model}$ for the data and model shown in Fig.~\ref{fig:distz}.  Blue dots show the weighted averages in bins with $N=204$ points for the high-redshift sample.  Weights were chosen to be $1/\sigma_\mu^2$, which $\sigma_\mu$ is the uncertainty of an individual measurement. Redshifts of the average    Red triangles show weighted averages in bins with $N= 100$ points per bin.  }
\label{fig:res1}
\end{figure}

\begin{figure}
\centering
\includegraphics[scale=0.5]{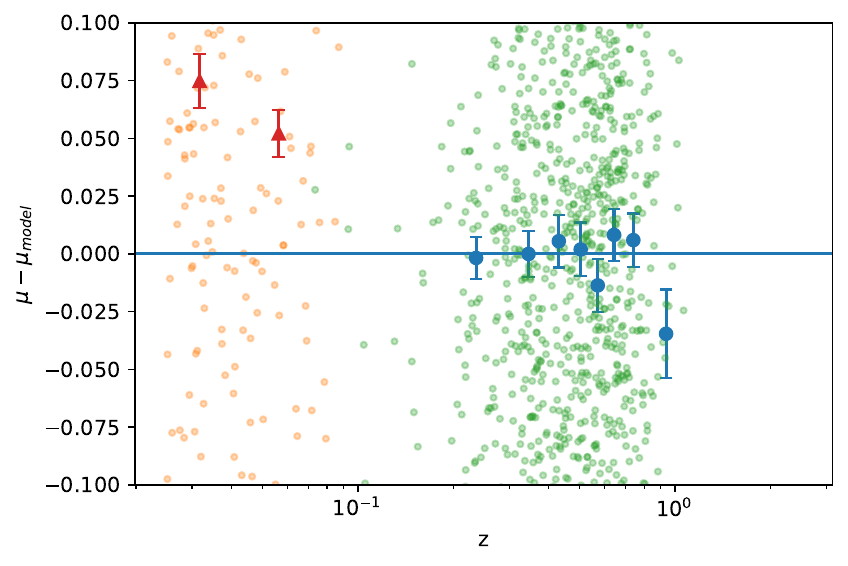}
\caption{Residuals $\mu -\mu_{model}$ for the data and model shown in Fig.~\ref{fig:distz} with the $y$-scale expanded to examine the residuals more closely.  Blue dots show the weighted averages in bins with $N=204$ points for the high-redshift sample.  Red triangles show averages in bins with $N= 100$ points per bin.  Error bars are the uncertainties in the averages calculated from the individual uncertainties of each point.  }
\label{fig:res2}
\end{figure}

We will consider the DES supernova data \citep{DES2024,DES2026} as consisting of two parts, the high redshift supernovae collected by the DES survey directly, and the low redshift sample which was collected by other surveys and re-analyzed by DES.   For simplicity we focus on the higher quality portion of the data, keeping only supernovae whose distance modulus uncertainty is less than 0.4 magnitudes, leaving us with 1,523 objects in our high redshift set and 197 in our low redshift set.  Now, since the DES sample does not have a calibrated zero point, the zero point must be marginalized over.  This is equivalent to marginalizing over the value of the Hubble constant.  In practice, constraints on $\Omega_m$ from the DES data are somewhat weaker than the constraints from the DESI data including the CMB constraint.  Given that the inclusion of the DES data into the analysis is unlikely to strengthen the result from DESI, we choose instead to examine consistency of the DES data with the best fit model from DESI.  First, we determine the value of $H_o$ that gives the best fit of the DES data to the DESI model using MCMC.  This comes out to be $H_o=70.495\pm 0.165$ km/s.  We remind the reader that this value is only applicable to this particular dataset and cannot be compared to other measurements of the Hubble constant.  In Fig.~\ref{fig:distz} we show the DES and DESI distance moduli vs. redshift together with the model using the best fit $\Omega_m$ from the DESI BAO and the best fit $H_o$ from the DES sample.  To show the overlap of the DES and DESI samples we show DES distance measurements as orange dots (Note that DES distances have been multiplied by $(1+z)$ to make them luminosity distances).  While this plot shows general good agreement with the model, it is important to examine the residuals between the data and the model.  Fig.~\ref{fig:res1} shows the residuals between the DES data and the model, along with averages in bins with equal numbers of objects (204 points per bin for the high redshift sample and 100 points per bin for the low redshift sample).  Fig.~\ref{fig:res2} is the same as Fig.~\ref{fig:res1} except the graph has been scaled to see more detail in the averages.  We see that the high redshift supernovae are consistent with the model.  Indeed, the $\chi^2$ per degree of freedom is $0.949$.  However, we also see that the low redshift supernovae differ significantly from the model in that the average of the residuals is shifted about 0.05 magnitudes in the positive direction.  This has been previously noted by \citet{Efs25}.

\begin{figure}
\centering
\includegraphics[scale=0.5]{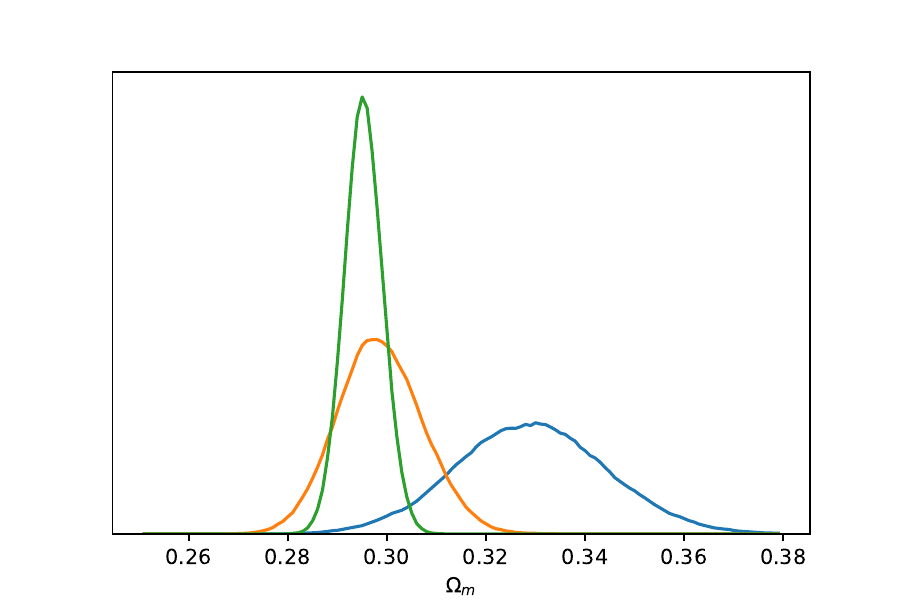}
\caption{Likelihoods for $\Omega_m$ for the entire sample (the blue curve) and when the low-redshift sample is removed (the orange curve).  For comparison we show the likelihood obtained from the DESI BAO data with the CMB constraint (green curve).  }
\label{fig:Omcomp}
\end{figure}

\begin{figure}
\centering
\includegraphics[scale=0.5]{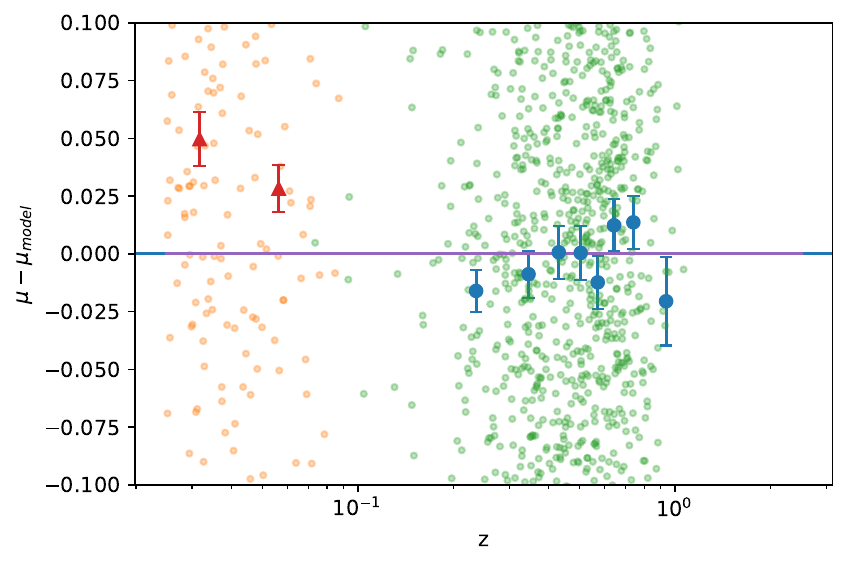}
\caption{Residuals $\mu -\mu_{model}$ for the DES data from the best fit model including the low redshift supernovae ($\Omega=0.33$ and $H_0=69.8$ km/s/Mpc).   Blue dots show the weighted averages in bins with $N=204$ points for the high-redshift sample.  Red triangles show averages in bins with $N= 100$ points per bin.  Error bars are the uncertainties in the averages calculated from the individual uncertainties of each point.}
\label{fig:reslgOm}
\end{figure}

It is said that data from the DES favors large values for $\Omega_m$.  This preference is almost entirely due to the deviation of the low redshift supernova from the best fit to the DESI BAO shown in Fig.~\ref{fig:res2}.  This is illustrated in Fig.~\ref{fig:Omcomp}, where we plot the likelihood of $\Omega_m$ for the entire sample (marginalizing over $H_o$) and compare it to the likelihood with the low redshift sample removed.   We see that, without the low-redshift sample, the likelihood of $\Omega_m$ for DES is quite consistent with that from DESI; however, including the low-redshift sample gives a measurement of $\Omega_m$ that is in some tension with the value obtained from the DESI BAO data with the CMB constraint.     Perhaps more importantly, even with the larger  $\Omega_m$ the model is still not a good fit to the entire sample.  In Fig~\ref{fig:reslgOm} we show the residuals for the best fit model to the entire sample.  We see that the low-redshift supernovae still have distance moduli that are systematically larger than expected.  
Indeed, it is the inability of the standard model to fit both the high and low redshift samples that is a large contributor to the preference of evolving dark energy over the standard model.  

\begin{figure}
\centering
\includegraphics[scale=0.5]{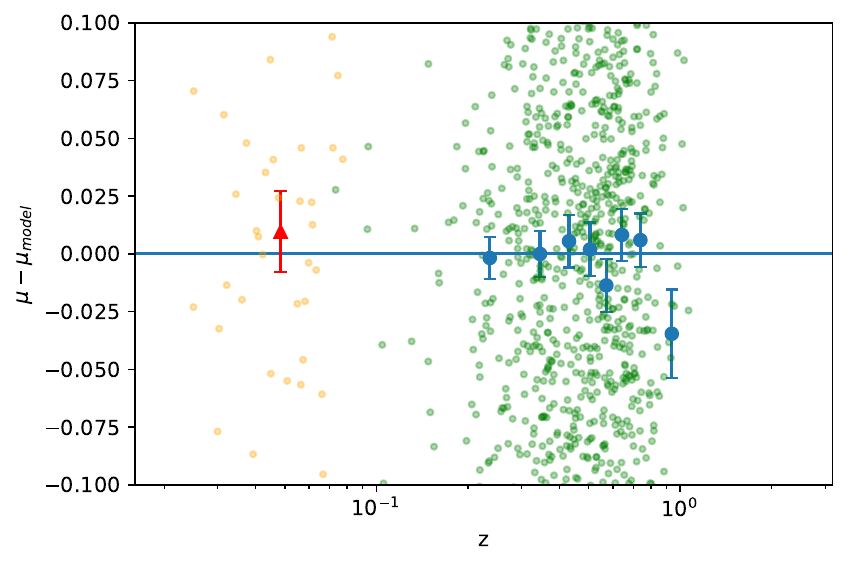}
\caption{The same as Fig.~\ref{fig:res2} except that the low-z supernovae from the DES have been replaced by those from DR0.5 from the DEBASS project.   }
\label{fig:resDEBASS}
\end{figure}

\cite{Efs25} has argued that the roughly 0.05 magnitude deviation of the low-z supernova from the standard cosmological model could be explained as a systematic, particularly since the supernova magnitudes were adjusted by shifts of the same order in adapting them to the DES system.  \citet{VinKesSha25} responded to this claim by justifying their analysis.  However, given the stakes riding on this small deviation, it would be preferable to have a low-z sample whose collection and analysis better matched the circumstances of the high-z sample.  Fortunately, the Dark Energy Bedrock All-Sky Supernova Program \citep[DEBASS,][]{SheAceBro26} is working on providing this low-z sample using the same telescope and analysis methods as the DES.  We have analyzed the preliminary data release DEBASS DR0.5, consisting of 55 low-z supernovae.  Fig.~\ref{fig:resDEBASS} is the same as Fig.~\ref{fig:res2} except that the low-z supernova sample from DES has been replaced with the DEBASS DR0.5.  Due to the small number of points we have calculated a single weighted average for the low-z sample.  From the figure we see that DEBASS sample is quite consistent with the standard cosmological model.  This suggests that when the DBASS replaces the current low-z sample in the DES the preference for the evolving dark energy model will drop significantly.  

\section{The Distance Redshift Relation}
\label{sec:DRR}

\begin{figure}
\centering
\includegraphics[scale=0.5]{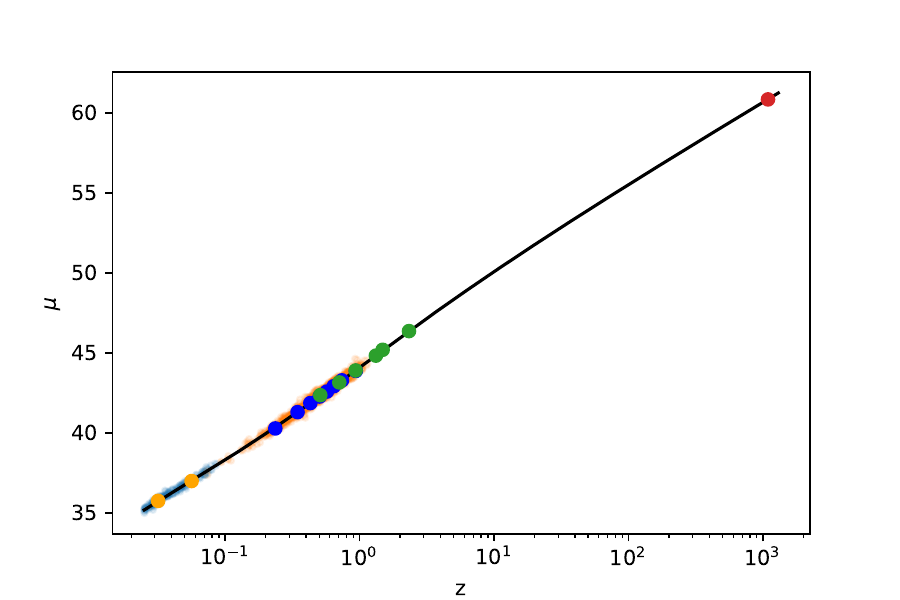}
\caption{The same as Fig.~\ref{fig:distz} except that we have included the acoustic peak of the cosmic background radiation as an additional bao at $z=1090$.  }
\label{fig:distred}
\end{figure}

If, for the moment, we assume that the low-z SN1a from the DEBASS project will continue to be consistent with the standard cosmological model described above, then we have an accurate model that describes  distance-redshift data from $z=0$ out to $z=1090$.  This is illustrated in Fig~\ref{fig:distred}, where we have again multiplied the BAO and CMB acoustic peak distances by $(1+z)$ to convert them to luminosity distances.  This is a significant confirmation of the standard cosmological model and overall an impressive achievement.  However, the idea that the standard cosmological model with a fixed value of $H_0$ can fit the distance-redshift relation from small $z$ all the way to the redshift at decoupling is difficult to reconcile with our current understanding of the Hubble tension.  

%Solutions to the Hubble tension generally involve either systematics in the data or changes to our cosmological model, such as new late time physics or changes to the circumstances around recombination.  However, if the standard model accurately models the expansion rate from the time of recombination to the present, this poses a real challenge to new physics explanations, since they would necessarily change the distance-redshift relation; indeed, the point of these models would be to interpolate between a low $H_0$ at high redshift to the larger ``local" value of the Hubble constant.  A successful distance-redshift modeling, then, strongly favors the Hubble tension being a result of systematics.   

Proposed solutions to the Hubble tension generally fall into two categories: unrecognized systematic errors in the data or modifications to the standard cosmological model, such as new late-time physics or changes to the physics of recombination. However, if the standard model accurately describes the expansion history from recombination to the present day, this presents a significant challenge for new-physics explanations, since such models necessarily alter the distance-redshift relation. Indeed, their purpose is to reconcile the low value of $H_0$ inferred at high redshift with the larger value measured locally. Consequently, strong observational agreement with the standard distance-redshift relation would strongly favors a resolution of the Hubble tension coming from systematic effects rather than new physics.

Recall that the SN1a and BAO data we are using do not by themselves allow us to determine the value of the Hubble constant $H_0$.  In the case of the supernova data, $H_0$ is ambiguous up to an unknown absolute magnitude.  The BAO data determines only the combination $h r_d$, where $H_0= 100h$ km/s/Mpc and $r_d$ the comoving sound horizon at the drag epoch.  In order to isolate $H_0$ we must add in some additional information.  

One approach is to calibrate the data using high-z information from the CMB \citep{CamDavHin25} .  This gives a constraint on $H_0$ that is consistent with that of Planck \citep{PlanckCosPar16}.  However, one could also use information from the local distance network that the Hubble constant for the SN1a should be $73.50\pm0.81$ km s$^{-1}$Mpc$^{-1}$  \citep{CanCruCso26}.  %do we want to discuss how this limits late time solutions to the Hubble tension???
Given that our model stretches continuously from low-z to high z with a constant $H_0$, we conclude that either information from the high-z CMB or the low-z local distance network must be incorrect.   Therefore there must be something about the Planck analysis or the local distance ladder that we don't understand.  

First we consider the local distance ladder.  Distances to SN1a are measured in terms of distance modulus, a logarithmic measure of distance.  When working with distance modulus, changing $H_0$ is equivalent to shifting all distance moduli by a constant, the same effect as changing the absolute magnitude zero point.  Making the Hubble constant of the SN1a match that from the CMB would involve shifting the zero point by 0.2 magnitudes.  However, the zero point is calibrated using SN1a that are in the same galaxies as another, more local, distance indicator, such as a Cepheid variable star or the Tip of the red giant branch.  These more local distance indicators are in turn calibrated from nearby instances where parallax is possible, the base rung of the distance ladder.  Much effort has been expended studying the systematics in this process and it has generally been found to be quite robust \citep[\eg,][]{ScoRieMur25}.  

Next we consider the CMB observations.  While here we have only used the angular size of the first acoustic peak,  there is much more information in the CMB, (nearly) all of which is exquisitely consistent with the cosmological standard model \citep{PlanckCosPar16,LouLaPAtk2025}.   It is difficult to imagine being able to maintain the agreement between the CMB data and our models using a larger value of the Hubble constant.  

\section{Discussion and Conclusions}
\label{sec:DnC}

We have examined the consistency of three of the most powerful probes of the cosmic distance scale: the CMB acoustic peak, DESI BAO measurements, and DES Type Ia supernova distances. Rather than performing a broad parameter-estimation exercise, our goal was to test whether a single flat $\Lambda$CDM model can simultaneously describe the distance-redshift relation inferred from these independent datasets.

Our first result shows that the DESI BAO \citep{DESI2025} measurements are in excellent agreement with the highly constrained parameter combination determined by the CMB \citep[\eg,][]{PlanckCosPar16}. The quantity $(r_d h)(\Omega_m/0.3)^{0.4}$, which is measured with extraordinary precision by Planck, is recovered by DESI at the sub-percent level. This agreement implies that a cosmological model passing through the acoustic-scale measurement at recombination also provides an excellent description of BAO distances at redshifts $z\sim1$. Given the vast range in cosmic time separating these measurements, this represents a striking confirmation of the standard cosmological framework.

Imposing the CMB constraint allows the DESI data to determine $\Omega_m$ with high precision, yielding a value somewhat lower than that preferred by Planck but still broadly consistent with the standard model. While this difference deserves continued attention, it is small compared with the discrepancies that arise when low-redshift supernova samples from DES are included in cosmological analyses.
Our second result is that the high-redshift DES \citep[\eg,][]{DES2024} supernova sample is fully consistent with the cosmological model favored by DESI and Planck. The residuals exhibit no discernible systematic trend, and the overall fit is exceptional. In contrast, the low-redshift supernova sample exhibits a coherent offset of approximately $0.05$ magnitudes relative to the model \citep[\eg,][]{Efs25}. This offset is responsible for much of the apparent preference for larger values of $\Omega_m$ and, more generally, for models with evolving dark energy. When the low-redshift sample is removed, the DES constraints become substantially more consistent with those derived from DESI.
The preliminary DEBASS sample \citep{SheAceBro26} provides an intriguing indication that this discrepancy may be attributable to systematic effects in the current low-redshift anchor sample. Although the present DEBASS release is too small to provide a definitive test, its supernova distances are consistent with the standard cosmological model and do not exhibit the offset seen in the low-redshift DES compilation. If this behavior persists in future releases, the motivation for departures from $\Lambda$CDM based on the DES+DESI+CMB distance measurements would be significantly weakened.

Taken together, these results suggest that the distance-redshift relation from the local Universe to recombination may already be described remarkably well by a single $\Lambda$CDM model. The combination of CMB, BAO, and supernova observations spans more than four orders of magnitude in redshift and probes the expansion history over nearly the entire age of the Universe. The fact that these measurements can be connected by a single smooth cosmological model is itself a powerful empirical result.
This conclusion has important implications for the Hubble tension \citep{Di_Valentino2021,CosmoVerse25}. Proposed resolutions generally invoke either previously unrecognized systematic errors or modifications to the standard cosmological model. However, any successful new-physics solution must alter the expansion history in a way that reconciles the lower value of $H_0$ inferred from high-redshift observations with the larger value measured locally. Such models therefore predict departures from the standard distance-redshift relation. If future observations continue to confirm the $\Lambda$CDM distance-redshift relation from low redshift through recombination, the available room for these modifications becomes increasingly restricted. In that case, the evidence would point more naturally toward unresolved systematic effects in one or more of the datasets used to infer $H_0$.

At present, the key question is not whether the standard cosmological model provides an adequate description of the observed distance scale, but whether the remaining discrepancies arise from subtle observational systematics or from new physics that preserves the remarkable agreement already seen across such a large range of redshifts. Future low-redshift supernova measurements, particularly from DEBASS and similar programs designed to match the observational and analysis framework of modern high-redshift surveys, will play a central role in answering that question.

\\
\\
\noindent{\bf Acknowledgements:} 
RW was partially supported by NSF grant AST-1907365.  HAF was partially supported by NSF grant AST-1907404.  
%Funding for the \textit{Cosmicflows} project has been provided by the US National Science Foundation grant AST09-08846, the National Aeronautics and Space Administration grant NNX12AE70G, and multiple awards to support observations with HST through the Space Telescope Science Institute.

%Data available on request. The data underlying this article will be shared on reasonable request to the corresponding author.

\bibliographystyle{mnras}
\bibliography{haf}

@article{Einstein1916,
    author = "Einstein, Albert",
    editor = "Hsu, Jong-Ping and Fine, D.",
    title = "{The foundation of the general theory of relativity.}",
    doi = "10.1002/andp.19163540702",
    journal = "Annalen Phys.",
    volume = "49",
    number = "7",
    pages = "769--822",
    year = "1916"
}

@article{Friedman1922,
    author = "Friedman, A.",
    title = "{On the Curvature of space}",
    doi = "10.1007/BF01332580",
    journal = "Z. Phys.",
    volume = "10",
    pages = "377--386",
    year = "1922"
}

@ARTICLE{Lemaitre1927,
       author = {{Lema{\^\i}tre}, G.},
        title = "{Un Univers homog{\`e}ne de masse constante et de rayon croissant rendant compte de la vitesse radiale des n{\'e}buleuses extra-galactiques}",
      journal = {Annales de la Soci{\'e}t{\'e} Scientifique de Bruxelles},
         year = 1927,
        month = jan,
       volume = {47},
        pages = {49-59},
       adsurl = {https://ui.adsabs.harvard.edu/abs/1927ASSB...47...49L}
}

@ARTICLE{ScoRieMur25,
       author = {{Scolnic}, Daniel and {Riess}, Adam G. and {Murakami}, Yukei S. and {Peterson}, Erik R. and {Brout}, Dillon and {Acevedo}, Maria and {Carreres}, Bastien and {Jones}, David O. and {Said}, Khaled and {Howlett}, Cullan and {Anand}, Gagandeep S.},
        title = "{The Hubble Tension in Our Own Backyard: DESI and the Nearness of the Coma Cluster}",
      journal = {\apjl},
         year = 2025,
        month = jan,
       volume = {979},
       number = {1},
          eid = {L9},
        pages = {L9},
          doi = {10.3847/2041-8213/ada0bd},
archivePrefix = {arXiv},
       eprint = {2409.14546},
 primaryClass = {astro-ph.CO},
       adsurl = {https://ui.adsabs.harvard.edu/abs/2025ApJ...979L...9S}
}

@ARTICLE{CanCruCso26,
       author = {{H0DN Collaboration} and {Casertano}, Stefano and {Anand}, Gagandeep and {Anderson}, Richard I. and {Beaton}, Rachael and {Bhardwaj}, Anupam and {Blakeslee}, John P. and {Boubel}, Paula and {Breuval}, Louise and {Brout}, Dillon and {Cantiello}, Michele and {Cruz Reyes}, Mauricio and {Cs{\"o}rnyei}, Geza and {de Jaeger}, Thomas and {Dhawan}, Suhail and {Di Valentino}, Eleonora and {Galbany}, Llu{\'\i}s and {Gil-Mar{\'\i}n}, H{\'e}ctor and {Graczyk}, Dariusz and {Huang}, Caroline and {Jensen}, Joseph B. and {Kervella}, Pierre and {Leibundgut}, Bruno and {Lengen}, Bastian and {Li}, Siyang and {Macri}, Lucas and {{\"O}z{\"u}lker}, Emre and {Pesce}, Dominic W. and {Riess}, Adam and {Romaniello}, Martino and {Said}, Khaled and {Sch{\"o}neberg}, Nils and {Scolnic}, Dan and {Sicignano}, Teresa and {Skowron}, Dorota M. and {Uddin}, Syed A. and {Verde}, Licia and {Nota}, Antonella},
        title = "{The Local Distance Network: A community consensus report on the measurement of the Hubble constant at {\ensuremath{\sim}}1\% precision}",
      journal = {\aap},
         year = 2026,
        month = apr,
       volume = {708},
          eid = {A166},
        pages = {A166},
          doi = {10.1051/0004-6361/202557993},
archivePrefix = {arXiv},
       eprint = {2510.23823},
 primaryClass = {astro-ph.CO},
       adsurl = {https://ui.adsabs.harvard.edu/abs/2026A&A...708A.166H}
}

@ARTICLE{SheAceBro26,
       author = {{Sherman}, Nora F. and {Acevedo}, Maria and {Brout}, Dillon and {Martin}, Bailey and {Scolnic}, Daniel and {Cao}, Dingyuan and {Lidman}, Christopher and {Ali}, Noor and {Armstrong}, Patrick and {Auchettl}, K. and {Chen}, Rebecca C. and {Drlica-Wagner}, Alex and {Ferguson}, Peter S. and {Herner}, Kenneth and {Narayan}, Gautham and {Peterson}, Erik R. and {Rauf}, Liana and {Rest}, Armin and {Riess}, Adam G. and {Sako}, Masao and {Schmidt}, Brian and {Tang}, Xianzhe TZ and {Tucker}, Brad E.},
        title = "{The Dark Energy Bedrock All-sky Supernova Program: Motivation, Design, Implementation, and Preliminary Data Release}",
      journal = {\apj},
         year = 2026,
        month = may,
       volume = {1002},
       number = {2},
          eid = {146},
        pages = {146},
          doi = {10.3847/1538-4357/ae48ff},
archivePrefix = {arXiv},
       eprint = {2508.10878},
 primaryClass = {astro-ph.CO},
       adsurl = {https://ui.adsabs.harvard.edu/abs/2026ApJ..1002..146S}
}

@ARTICLE{CamDavHin25,
       author = {{Camilleri}, R. and {Davis}, T.~M. and {Hinton}, S.~R. and {Armstrong}, P. and {Brout}, D. and {Galbany}, L. and {Glazebrook}, K. and {Lee}, J. and {Lidman}, C. and {M{\"o}ller}, A. and {Nichol}, R.~C. and {Sako}, M. and {Scolnic}, D. and {Shah}, P. and {Smith}, M. and {Sullivan}, M. and {S{\'a}nchez}, B.~O. and {Vincenzi}, M. and {Wiseman}, P. and {Allam}, S. and {Abbott}, T.~M.~C. and {Aguena}, M. and {Andrade-Oliveira}, F. and {Asorey}, J. and {Avila}, S. and {Bacon}, D. and {Bechtol}, K. and {Bocquet}, S. and {Brooks}, D. and {Buckley-Geer}, E. and {Burke}, D.~L. and {Rosell}, A. Carnero and {Carollo}, D. and {Carretero}, J. and {Castander}, F.~J. and {Conselice}, C. and {da Costa}, L.~N. and {Pereira}, M.~E.~S. and {Desai}, S. and {Diehl}, H.~T. and {Everett}, S. and {Ferrero}, I. and {Flaugher}, B. and {Frieman}, J. and {Garc{\'\i}a-Bellido}, J. and {Gaztanaga}, E. and {Giannini}, G. and {Gruendl}, R.~A. and {Herner}, K. and {Hollowood}, D.~L. and {Honscheid}, K. and {Huterer}, D. and {James}, D.~J. and {Kent}, S. and {Kuehn}, K. and {Lahav}, O. and {Lee}, S. and {Lewis}, G.~F. and {Lima}, M. and {Marshall}, J.~L. and {Mena-Fern{\'a}ndez}, J. and {Miquel}, R. and {Myles}, J. and {Ogando}, R.~L.~C. and {Palmese}, A. and {Pieres}, A. and {Plazas Malag{\'o}n}, A.~A. and {Romer}, A.~K. and {Roodman}, A. and {Samuroff}, S. and {Sanchez}, E. and {Sanchez Cid}, D. and {Schubnell}, M. and {Sevilla-Noarbe}, I. and {Suchyta}, E. and {Suntzeff}, N. and {Swanson}, M.~E.~C. and {Tarle}, G. and {Tucker}, B.~E. and {Walker}, A.~R. and {Weaverdyck}, N. and {DES Collaboration}},
        title = "{The Dark Energy Survey Supernova Program: an updated measurement of the Hubble constant using the inverse distance ladder}",
      journal = {\mnras},
         year = 2025,
        month = feb,
       volume = {537},
       number = {2},
        pages = {1818-1825},
          doi = {10.1093/mnras/staf122},
archivePrefix = {arXiv},
       eprint = {2406.05049},
 primaryClass = {astro-ph.CO},
       adsurl = {https://ui.adsabs.harvard.edu/abs/2025MNRAS.537.1818C}
}

@article{PopShaKen2026,
    author = {Popovic, B and Shah, P and Kenworthy, W D and Kessler, R and Davis, T M and Goobar, A and Scolnic, D and Vincenzi, M and Wiseman, P and Chen, R and Charleton, E and Acevedo, M and Armstrong, P and Boyd, B M and Brout, D and Camilleri, R and Frieman, J and Galbany, L and Grayling, M and Kelsey, L and Rose, B and Sánchez, B and Lee, J and Möller, A and Smith, M and Sullivan, M and Shiamtanis, N and Alarcon, A and Allam, S S and Andrade-Oliveira, F and Avila, S and Bacon, D and Blazek, J and Bocquet, S and Brooks, D and Burke, D L and Rosell, A Carnero and Carretero, J and Cawthon, R and da Costa, L N and Pereira, M E da Silva and Diehl, H T and Dodelson, S and Doel, P and Everett, S and Frohmaier, C and García-Bellido, J and Gruen, D and Gutierrez, G and Herner, K and Hinton, S R and Hollowood, D L and Honscheid, K and Huterer, D and James, D J and Jeffrey, N and Kuehn, K and Lahav, O and Lee, S and Lidman, C and Marshall, J L and Mena-Fernández, J and Menanteau, F and Miquel, R and Muir, J and Myles, J and Ogando, R L C and Paterno, M and Malagón, A A Plazas and Porredon, A and Prat, J and Nichol, R C and Romer, A K and Roodman, A and Sanchez, E and Cid, D Sanchez and Sevilla-Noarbe, I and Suchyta, E and Swanson, M E C and To, C and Tucker, D L and Walker, A R and Weaverdyck, N and Aguena, M, The DES Collaboration},
    title = {The Dark Energy Survey supernova program: a reanalysis of cosmology results and evidence for evolving dark energy with an updated Type Ia supernova calibration},
    journal = {Monthly Notices of the Royal Astronomical Society},
    volume = {548},
    number = {4},
    pages = {stag632},
    year = {2026},
    month = {06},
    issn = {0035-8711},
    doi = {10.1093/mnras/stag632},
    url = {https://doi.org/10.1093/mnras/stag632},
    eprint = {https://academic.oup.com/mnras/article-pdf/548/4/stag632/68079698/stag632.pdf},
}

@ARTICLE{AceSheBro2026,
       author = {{Acevedo}, Maria and {Sherman}, Nora F. and {Brout}, Dillon and {Carreres}, Bastien and {Scolnic}, Daniel and {Popovic}, Brodie and {Armstrong}, Patrick and {Cao}, Dingyuan and {Chen}, Rebecca C. and {Drlica-Wagner}, Alex and {Ferguson}, Peter S. and {Lidman}, Christopher and {Martin}, Bailey and {Peterson}, Erik R. and {Riess}, Adam G.},
        title = "{The Dark Energy Bedrock All-sky Supernova Program: Cross Calibration, Simulations, and Cosmology Forecasts}",
      journal = {\apj},
         year = 2026,
        month = jan,
       volume = {996},
       number = {1},
          eid = {7},
        pages = {7},
          doi = {10.3847/1538-4357/ae1e78},
archivePrefix = {arXiv},
       eprint = {2508.10877},
 primaryClass = {astro-ph.CO},
       adsurl = {https://ui.adsabs.harvard.edu/abs/2026ApJ...996....7A}
}

@ARTICLE{SheAveBro2026,
       author = {{Sherman}, Nora F. and {Acevedo}, Maria and {Brout}, Dillon and {Martin}, Bailey and {Scolnic}, Daniel and {Cao}, Dingyuan and {Lidman}, Christopher and {Ali}, Noor and {Armstrong}, Patrick and {Auchettl}, K. and {Chen}, Rebecca C. and {Drlica-Wagner}, Alex and {Ferguson}, Peter S. and {Herner}, Kenneth and {Narayan}, Gautham and {Peterson}, Erik R. and {Rauf}, Liana and {Rest}, Armin and {Riess}, Adam G. and {Sako}, Masao and {Schmidt}, Brian and {Tang}, Xianzhe TZ and {Tucker}, Brad E.},
        title = "{The Dark Energy Bedrock All-sky Supernova Program: Motivation, Design, Implementation, and Preliminary Data Release}",
      journal = {\apj},
         year = 2026,
        month = may,
       volume = {1002},
       number = {2},
          eid = {146},
        pages = {146},
          doi = {10.3847/1538-4357/ae48ff},
archivePrefix = {arXiv},
       eprint = {2508.10878},
 primaryClass = {astro-ph.CO},
       adsurl = {https://ui.adsabs.harvard.edu/abs/2026ApJ..1002..146S}
}

@ARTICLE{VinKesSha25,
       author = {{Vincenzi}, M. and {Kessler}, R. and {Shah}, P. and {Lee}, J. and {Davis}, T.~M. and {Scolnic}, D. and {Armstrong}, P. and {Brout}, D. and {Camilleri}, R. and {Chen}, R. and {Galbany}, L. and {Lidman}, C. and {M{\"o}ller}, A. and {Popovic}, B. and {Rose}, B. and {Sako}, M. and {S{\'a}nchez}, B.~O. and {Smith}, M. and {Sullivan}, M. and {Wiseman}, P. and {Abbott}, T.~M.~C. and {Aguena}, M. and {Allam}, S. and {Andrade-Oliveira}, F. and {Bocquet}, S. and {Brooks}, D. and {Carnero Rosell}, A. and {Carretero}, J. and {da Costa}, L.~N. and {Pereira}, M.~E.~S. and {Diehl}, H.~T. and {Doel}, P. and {Everett}, S. and {Flaugher}, B. and {Frieman}, J. and {Garc{\'\i}a-Bellido}, J. and {Gaztanaga}, E. and {Gruen}, D. and {Gruendl}, R.~A. and {Gutierrez}, G. and {Hinton}, S.~R. and {Hollowood}, D.~L. and {Honscheid}, K. and {James}, D.~J. and {Kuehn}, K. and {Lahav}, O. and {Lee}, S. and {Marshall}, J.~L. and {Mena-Fern{\'a}ndez}, J. and {Miquel}, R. and {Muir}, J. and {Myles}, J. and {Palmese}, A. and {Plazas Malag{\'o}n}, A.~A. and {Porredon}, A. and {Samuroff}, S. and {Sanchez}, E. and {Sanchez Cid}, D. and {Sevilla-Noarbe}, I. and {Suchyta}, E. and {Tarle}, G. and {To}, C. and {Tucker}, D.~L. and {Vikram}, V. and {Walker}, A.~R. and {Weaverdyck}, N. and {Weller}, J.},
        title = "{Comparing the DES-SN5YR and Pantheon+ SN cosmology analyses: investigation based on 'evolving dark energy or supernovae systematics'?}",
      journal = {\mnras},
         year = 2025,
        month = aug,
       volume = {541},
       number = {3},
        pages = {2585-2593},
          doi = {10.1093/mnras/staf943},
archivePrefix = {arXiv},
       eprint = {2501.06664},
 primaryClass = {astro-ph.CO},
       adsurl = {https://ui.adsabs.harvard.edu/abs/2025MNRAS.541.2585V}
}

@ARTICLE{GiaAladas2025,
       author = {{Giannini}, G. and {Alarcon}, A. and {d'Assignies}, W. and {Bernstein}, G.~M. and {Troxel}, M.~A. and {Chang}, C. and {Yin}, B. and {Amon}, A. and {Myles}, J. and {Weaverdyck}, N. and {Porredon}, A. and {Anbajagane}, D. and {Avila}, S. and {Bechtol}, K. and {Becker}, M.~R. and {Blazek}, J. and {Crocce}, M. and {Gruen}, D. and {Rodriguez-Monroy}, M. and {S{\'a}nchez}, C. and {Sanchez Cid}, D. and {Sevilla-Noarbe}, I. and {Aguena}, M. and {Allam}, S. and {Alves}, O. and {Andrade-Oliveira}, F. and {Bacon}, D. and {Bocquet}, S. and {Brooks}, D. and {Camilleri}, R. and {Carnero Rosell}, A. and {Carretero}, J. and {Cawthon}, R. and {da Costa}, L.~N. and {da Silva Pereira}, M.~E. and {Davis}, T.~M. and {De Vicente}, J. and {DePoy}, D.~L. and {Desai}, S. and {Diehl}, H.~T. and {Dodelson}, S. and {Doel}, P. and {Doux}, C. and {Drlica-Wagner}, A. and {Elvin-Poole}, J. and {Everett}, S. and {Evrard}, A.~E. and {Flaugher}, B. and {Frieman}, J. and {Garc{\'\i}a-Bellido}, J. and {Gatti}, M. and {Gaztanaga}, E. and {Giles}, P. and {Gruendl}, R.~A. and {Gutierrez}, G. and {Herner}, K. and {Hinton}, S.~R. and {Hollowood}, D.~L. and {Honscheid}, K. and {Huterer}, D. and {James}, D.~J. and {Kuehn}, K. and {Lahav}, O. and {Lee}, S. and {Lin}, H. and {Marshall}, J.~L. and {Mena-Fern{\'a}ndez}, J. and {Menanteau}, F. and {Miquel}, R. and {Muir}, J. and {Ogando}, R.~L.~C. and {Petravick}, D. and {Plazas Malag{\'o}n}, A.~A. and {Prat}, J. and {Raveri}, M. and {Rykoff}, E.~S. and {Samuroff}, S. and {Sanchez}, E. and {Shin}, T. and {Smith}, M. and {Suchyta}, E. and {Swanson}, M.~E.~C. and {Tarle}, G. and {Thomas}, D. and {To}, C. and {Tucker}, D.~L. and {Vikram}, V. and {Yamamoto}, M.},
        title = "{Dark Energy Survey Year 6 Results: Redshift Calibration of the MagLim++ Lens Sample}",
      journal = {arXiv e-prints},
         year = 2025,
        month = sep,
          eid = {arXiv:2509.07964},
        pages = {arXiv:2509.07964},
          doi = {10.48550/arXiv.2509.07964},
archivePrefix = {arXiv},
       eprint = {2509.07964},
 primaryClass = {astro-ph.CO},
       adsurl = {https://ui.adsabs.harvard.edu/abs/2025arXiv250907964G}
}

@ARTICLE{LouLaPAtk2025,
       author = {{Louis}, Thibaut and {La Posta}, Adrien and {Atkins}, Zachary and {Jense}, Hidde T. and {Abril-Cabezas}, Irene and {Addison}, Graeme E. and {Ade}, Peter A.~R. and {Aiola}, Simone and {Alford}, Tommy and {Alonso}, David and {Amiri}, Mandana and {An}, Rui and {Austermann}, Jason E. and {Barbavara}, Eleonora and {Battaglia}, Nicholas and {Battistelli}, Elia Stefano and {Beall}, James A. and {Bean}, Rachel and {Beheshti}, Ali and {Beringue}, Benjamin and {Bhandarkar}, Tanay and {Biermann}, Emily and {Bolliet}, Boris and {Bond}, J. Richard and {Calabrese}, Erminia and {Capalbo}, Valentina and {Carrero}, Felipe and {Chen}, Shi-Fan and {Chesmore}, Grace and {Cho}, Hsiao-mei and {Choi}, Steve K. and {Clark}, Susan E. and {Cothard}, Nicholas F. and {Coughlin}, Kevin and {Coulton}, William and {Crichton}, Devin and {Crowley}, Kevin T. and {Darwish}, Omar and {Devlin}, Mark J. and {Dicker}, Simon and {Duell}, Cody J. and {Duff}, Shannon M. and {Duivenvoorden}, Adriaan J. and {Dunkley}, Jo and {Dunner}, Rolando and {Embil Villagra}, Carmen and {Fankhanel}, Max and {Farren}, Gerrit S. and {Ferraro}, Simone and {Foster}, Allen and {Freundt}, Rodrigo and {Fuzia}, Brittany and {Gallardo}, Patricio A. and {Garrido}, Xavier and {Gerbino}, Martina and {Giardiello}, Serena and {Gill}, Ajay and {Givans}, Jahmour and {Gluscevic}, Vera and {Goldstein}, Samuel and {Golec}, Joseph E. and {Gong}, Yulin and {Guan}, Yilun and {Halpern}, Mark and {Harrison}, Ian and {Hasselfield}, Matthew and {Healy}, Erin and {Henderson}, Shawn and {Hensley}, Brandon and {Herv{\'\i}as-Caimapo}, Carlos and {Hill}, J. Colin and {Hilton}, Gene C. and {Hilton}, Matt and {Hincks}, Adam D. and {Hlo{\v{z}}ek}, Ren{\'e}e and {Ho}, Shuay-Pwu Patty and {Hood}, John and {Hornecker}, Erika and {Huber}, Zachary B. and {Hubmayr}, Johannes and {Huffenberger}, Kevin M. and {Hughes}, John P. and {Ikape}, Margaret and {Irwin}, Kent and {Isopi}, Giovanni and {Joshi}, Neha and {Keller}, Ben and {Kim}, Joshua and {Knowles}, Kenda and {Koopman}, Brian J. and {Kosowsky}, Arthur and {Kramer}, Darby and {Kusiak}, Aleksandra and {Lagu{\"e}}, Alex and {Lakey}, Victoria and {Lee}, Eunseong and {Li}, Yaqiong and {Li}, Zack and {Limon}, Michele and {Lokken}, Martine and {Lungu}, Marius and {MacCrann}, Niall and {MacInnis}, Amanda and {Madhavacheril}, Mathew S. and {Maldonado}, Diego and {Maldonado}, Felipe and {Mallaby-Kay}, Maya and {Marques}, Gabriela A. and {van Marrewijk}, Joshiwa and {McCarthy}, Fiona and {McMahon}, Jeff and {Mehta}, Yogesh and {Menanteau}, Felipe and {Moodley}, Kavilan and {Morris}, Thomas W. and {Mroczkowski}, Tony and {Naess}, Sigurd and {Namikawa}, Toshiya and {Nati}, Federico and {Nerval}, Simran K. and {Newburgh}, Laura and {Nicola}, Andrina and {Niemack}, Michael D. and {Nolta}, Michael R. and {Orlowski-Scherer}, John and {Pagano}, Luca and {Page}, Lyman A. and {Pandey}, Shivam and {Partridge}, Bruce and {Perez Sarmiento}, Karen and {Prince}, Heather and {Puddu}, Roberto and {Qu}, Frank J. and {Ragavan}, Damien C. and {Ried Guachalla}, Bernardita and {Rogers}, Keir K. and {Rojas}, Felipe and {Sakuma}, Tai and {Schaan}, Emmanuel and {Schmitt}, Benjamin L. and {Sehgal}, Neelima and {Shaikh}, Shabbir and {Sherwin}, Blake D. and {Sierra}, Carlos and {Sievers}, Jon and {Sif{\'o}n}, Crist{\'o}bal and {Simon}, Sara and {Sonka}, Rita and {Spergel}, David N. and {Staggs}, Suzanne T. and {Storer}, Emilie and {Surrao}, Kristen and {Switzer}, Eric R. and {Tampier}, Niklas and {Thornton}, Robert and {Trac}, Hy and {Tucker}, Carole and {Ullom}, Joel and {Vale}, Leila R. and {Van Engelen}, Alexander and {Van Lanen}, Jeff and {Vargas}, Cristian and {Vavagiakis}, Eve M. and {Wagoner}, Kasey and {Wang}, Yuhan and {Wenzl}, Lukas and {Wollack}, Edward J. and {Zheng}, Kaiwen and {The Atacama Cosmology Telescope collaboration}},
        title = "{The Atacama Cosmology Telescope: DR6 power spectra, likelihoods and {\ensuremath{\Lambda}}CDM parameters}",
      journal = {\jcap},
         year = 2025,
        month = nov,
       volume = {2025},
       number = {11},
          eid = {062},
        pages = {062},
          doi = {10.1088/1475-7516/2025/11/062},
archivePrefix = {arXiv},
       eprint = {2503.14452},
 primaryClass = {astro-ph.CO},
       adsurl = {https://ui.adsabs.harvard.edu/abs/2025JCAP...11..062L}
}

@ARTICLE{Efs25b,
       author = {{Efstathiou}, George},
        title = "{Baryon acoustic oscillations from a different angle}",
      journal = {\mnras},
         year = 2025,
        month = jul,
       volume = {540},
       number = {3},
        pages = {2844-2852},
          doi = {10.1093/mnras/staf906},
archivePrefix = {arXiv},
       eprint = {2505.02658},
 primaryClass = {astro-ph.CO},
       adsurl = {https://ui.adsabs.harvard.edu/abs/2025MNRAS.540.2844E}
}

@ARTICLE{Efs25,
       author = {{Efstathiou}, George},
        title = "{Evolving dark energy or supernovae systematics?}",
      journal = {\mnras},
         year = 2025,
        month = apr,
       volume = {538},
       number = {2},
        pages = {875-882},
          doi = {10.1093/mnras/staf301},
archivePrefix = {arXiv},
       eprint = {2408.07175},
 primaryClass = {astro-ph.CO},
       adsurl = {https://ui.adsabs.harvard.edu/abs/2025MNRAS.538..875E}
}

@ARTICLE{SaiColMag20,
       author = {{Said}, Khaled and {Colless}, Matthew and {Magoulas}, Christina and {Lucey}, John R. and {Hudson}, Michael J.},
        title = "{Joint analysis of 6dFGS and SDSS peculiar velocities for the growth rate of cosmic structure and tests of gravity}",
      journal = {\mnras},
         year = 2020,
        month = sep,
       volume = {497},
       number = {1},
        pages = {1275-1293},
          doi = {10.1093/mnras/staa2032},
archivePrefix = {arXiv},
       eprint = {2007.04993},
 primaryClass = {astro-ph.CO},
       adsurl = {https://ui.adsabs.harvard.edu/abs/2020MNRAS.497.1275S}
}

@ARTICLE{CF4-full,
       author = {{Tully}, R. Brent and {Kourkchi}, Ehsan and {Courtois}, H{\'e}l{\`e}ne M. and {Anand}, Gagandeep S. and {Blakeslee}, John P. and {Brout}, Dillon and {Jaeger}, Thomas de and {Dupuy}, Alexandra and {Guinet}, Daniel and {Howlett}, Cullan and {Jensen}, Joseph B. and {Pomar{\`e}de}, Daniel and {Rizzi}, Luca and {Rubin}, David and {Said}, Khaled and {Scolnic}, Daniel and {Stahl}, Benjamin E.},
        title = "{Cosmicflows-4}",
      journal = {\apj},
         year = 2023,
        month = feb,
       volume = {944},
       number = {1},
          eid = {94},
        pages = {94},
          doi = {10.3847/1538-4357/ac94d8},
archivePrefix = {arXiv},
       eprint = {2209.11238},
 primaryClass = {astro-ph.CO},
       adsurl = {https://ui.adsabs.harvard.edu/abs/2023ApJ...944...94T}
}

@ARTICLE{RieYuaMac22,
       author = {{Riess}, Adam G. and {Yuan}, Wenlong and {Macri}, Lucas M. and {Scolnic}, Dan and {Brout}, Dillon and {Casertano}, Stefano and {Jones}, David O. and {Murakami}, Yukei and {Anand}, Gagandeep S. and {Breuval}, Louise and {Brink}, Thomas G. and {Filippenko}, Alexei V. and {Hoffmann}, Samantha and {Jha}, Saurabh W. and {D'arcy Kenworthy}, W. and {Mackenty}, John and {Stahl}, Benjamin E. and {Zheng}, WeiKang},
        title = "{A Comprehensive Measurement of the Local Value of the Hubble Constant with 1 km s$^{-1}$ Mpc$^{-1}$ Uncertainty from the Hubble Space Telescope and the SH0ES Team}",
      journal = {\apjl},
         year = 2022,
        month = jul,
       volume = {934},
       number = {1},
          eid = {L7},
        pages = {L7},
          doi = {10.3847/2041-8213/ac5c5b},
archivePrefix = {arXiv},
       eprint = {2112.04510},
 primaryClass = {astro-ph.CO},
       adsurl = {https://ui.adsabs.harvard.edu/abs/2022ApJ...934L...7R}
}

@ARTICLE{HowlettSDSS,
       author = {{Howlett}, Cullan and {Said}, Khaled and {Lucey}, John R. and {Colless}, Matthew and {Qin}, Fei and {Lai}, Yan and {Tully}, R. Brent and {Davis}, Tamara M.},
        title = "{The Sloan Digital Sky Survey peculiar velocity catalogue}",
      journal = {\mnras},
         year = 2022,
        month = sep,
       volume = {515},
       number = {1},
        pages = {953-976},
          doi = {10.1093/mnras/stac1681},
archivePrefix = {arXiv},
       eprint = {2201.03112},
 primaryClass = {astro-ph.CO},
       adsurl = {https://ui.adsabs.harvard.edu/abs/2022MNRAS.515..953H}
}

@ARTICLE{CosmoVerse25,
       author = {{Di Valentino}, Eleonora and {Said}, Jackson Levi and {Riess}, Adam and {Pollo}, Agnieszka and {Poulin}, Vivian and {G{\'o}mez-Valent}, Adri{\`a} and {Weltman}, Amanda and {Palmese}, Antonella and {Huang}, Caroline D. and {van de Bruck}, Carsten and {Saraf}, Chandra Shekhar and {Kuo}, Cheng-Yu and {Uhlemann}, Cora and {Grand{\'o}n}, Daniela and {Paz}, Dante and {Eckert}, Dominique and {Teixeira}, Elsa M. and {Saridakis}, Emmanuel N. and {Colg{\'a}in}, Eoin {\'O}. and {Beutler}, Florian and {Niedermann}, Florian and {Bajardi}, Francesco and {Barenboim}, Gabriela and {Gubitosi}, Giulia and {Musella}, Ilaria and {Banik}, Indranil and {Szapudi}, Istvan and {Singal}, Jack and {Cases}, Jaume Haro and {Chluba}, Jens and {Torrado}, Jes{\'u}s and {Mifsud}, Jurgen and {Jedamzik}, Karsten and {Said}, Khaled and {Dialektopoulos}, Konstantinos and {Herold}, Laura and {Perivolaropoulos}, Leandros and {Zu}, Lei and {Galbany}, Llu{\'\i}s and {Breuval}, Louise and {Visinelli}, Luca and {Escamilla}, Luis A. and {Anchordoqui}, Luis A. and {Sheikh-Jabbari}, M.~M. and {Lembo}, Margherita and {Dainotti}, Maria Giovanna and {Vincenzi}, Maria and {Asgari}, Marika and {Gerbino}, Martina and {Forconi}, Matteo and {Cantiello}, Michele and {Moresco}, Michele and {Benetti}, Micol and {Sch{\"o}neberg}, Nils and {Akarsu}, {\"O}zg{\"u}r and {Nunes}, Rafael C. and {Bernardo}, Reginald Christian and {Ch{\'a}vez}, Ricardo and {Anderson}, Richard I. and {Watkins}, Richard and {Capozziello}, Salvatore and {Li}, Siyang and {Vagnozzi}, Sunny and {Pan}, Supriya and {Treu}, Tommaso and {Irsic}, Vid and {Handley}, Will and {Giar{\`e}}, William and {Murakami}, Yukei and {Banihashemi}, Abdolali and {Poudou}, Ad{\`e}le and {Heavens}, Alan and {Kogut}, Alan and {Domi}, Alba and {Lenart}, Aleksander {\L}ukasz and {Melchiorri}, Alessandro and {Vadal{\`a}}, Alessandro and {Amon}, Alexandra and {Rivera}, Alexander Bonilla and {Reeves}, Alexander and {Zhuk}, Alexander and {Bonanno}, Alfio and {{\"O}vg{\"u}n}, Ali and {Pisani}, Alice and {Talebian}, Alireza and {Abebe}, Amare and {Aboubrahim}, Amin and {Gonz{\'a}lez Mor{\'a}n}, Ana Luisa and {Kov{\'a}cs}, Andr{\'a}s and {Lymperis}, Andreas and {Papatriantafyllou}, Andreas and {Liddle}, Andrew R. and {Paliathanasis}, Andronikos and {Borowiec}, Andrzej and {Yadav}, Anil Kumar and {Yadav}, Anita and {Sen}, Anjan Ananda and {William}, Anjitha John and {Davis}, Anne Christine and {Shajib}, Anowar J. and {Walters}, Anthony and {Lonappan}, Anto Idicherian and {Chudaykin}, Anton and {Capodagli}, Antonio and {da Silva}, Antonio and {De Felice}, Antonio and {Racioppi}, Antonio and {Oficial}, Araceli Soler and {Montiel}, Ariadna and {Favale}, Arianna and {Bernui}, Armando and {Velasco}, Arrianne Crystal and {Heinesen}, Asta and {Bakopoulos}, Athanasios and {Chatzistavrakidis}, Athanasios and {Khanpour}, Bahman and {Sathyaprakash}, Bangalore S. and {Zgirski}, Bartek and {L'Huillier}, Benjamin and {Famaey}, Benoit and {Jain}, Bhuvnesh and {Zhang}, Bing and {Karmakar}, Biswajit and {Dragovich}, Branko and {Thomas}, Brooks and {Correa}, Carlos and {Boiza}, Carlos G. and {Marques}, Catarina and {Escamilla-Rivera}, Celia and {Tzerefos}, Charalampos and {Zhang}, Chi and {De Leo}, Chiara and {Pfeifer}, Christian and {Lee}, Christine and {Venter}, Christo and {Gomes}, Cl{\'a}udio and {Roque De bom}, Clecio and {Moreno-Pulido}, Cristian and {Iosifidis}, Damianos and {Grin}, Dan and {Blixt}, Daniel and {Scolnic}, Dan and {Oriti}, Daniele and {Dobrycheva}, Daria and {Bettoni}, Dario and {Benisty}, David and {Fern{\'a}ndez-Arenas}, David and {Wiltshire}, David L. and {Sanchez Cid}, David and {Tamayo}, David and {Valls-Gabaud}, David and {Pedrotti}, Davide and {Wang}, Deng and {Staicova}, Denitsa and {Totolou}, Despoina and {Rubiera-Garcia}, Diego and {Milakovi{\'c}}, Dinko and {Pesce}, Dominic W. and {Sluse}, Dominique and {Borka}, Du{\v{s}}ko and {Yusofi}, Ebrahim and {Giusarma}, Elena and {Terlevich}, Elena and {Tomasetti}, Elena and {Vagenas}, Elias C. and {Fazzari}, Elisa and {Ferreira}, Elisa G.~M. and {Barakovic}, Elvis and {Dimastrogiovanni}, Emanuela and {Holm}, Emil Brinch and {Mottola}, Emil and {{\"O}z{\"u}lker}, Emre and {Specogna}, Enrico and {Brocato}, Enzo and {Jensko}, Erik and {Enriquez}, Erika Antonette and {Bhatia}, Esha and {Bresolin}, Fabio and {Avila}, Felipe and {Bouch{\`e}}, Filippo and {Bombacigno}, Flavio and {Anagnostopoulos}, Fotios K. and {Pace}, Francesco and {Sorrenti}, Francesco and {Lobo}, Francisco S.~N. and {Courbin}, Fr{\'e}d{\'e}ric and {Hansen}, Frode K. and {Sloan}, Greg and {Farrugia}, Gabriel and {Lynch}, Gabriel and {Garcia-Arroyo}, Gabriela and {Raimondo}, Gabriella and {Lambiase}, Gaetano and {Anand}, Gagandeep S. and {Poulot}, Gaspard and {Leon}, Genly and {Kouniatalis}, Gerasimos and {Nardini}, Germano and {Cs{\"o}rnyei}, G{\'e}za and {Galloni}, Giacomo},
        title = "{The CosmoVerse White Paper: Addressing observational tensions in cosmology with systematics and fundamental physics}",
      journal = {Physics of the Dark Universe},
         year = 2025,
        month = sep,
       volume = {49},
          eid = {101965},
        pages = {101965},
          doi = {10.1016/j.dark.2025.101965},
archivePrefix = {arXiv},
       eprint = {2504.01669},
 primaryClass = {astro-ph.CO},
       adsurl = {https://ui.adsabs.harvard.edu/abs/2025PDU....4901965D}
}

@misc{HuWang2023,
      title={Hubble Tension: The Evidence of New Physics}, 
      author={Jian-Ping Hu and Fa-Yin Wang},
      year={2023},
      eprint={2302.05709},
      archivePrefix={arXiv},
      primaryClass={astro-ph.CO},
      url={https://arxiv.org/abs/2302.05709}, 
}

@ARTICLE{DaiDeSSch2021,
       author = {{Dainotti}, M.~G. and {De Simone}, B. and {Schiavone}, T. and {Montani}, G. and {Rinaldi}, E. and {Lambiase}, G.},
        title = "{On the Hubble Constant Tension in the SNe Ia Pantheon Sample}",
      journal = {\apj},
         year = 2021,
        month = may,
       volume = {912},
       number = {2},
          eid = {150},
        pages = {150},
          doi = {10.3847/1538-4357/abeb73},
archivePrefix = {arXiv},
       eprint = {2103.02117},
 primaryClass = {astro-ph.CO},
       adsurl = {https://ui.adsabs.harvard.edu/abs/2021ApJ...912..150D}
}

@article{Di_Valentino2021,
	doi = {10.1088/1361-6382/ac086d}, 
	url = {https://doi.org/10.1088%2F1361-6382%2Fac086d},
  	year = 2021,
	month = {jul},
  	publisher = {{IOP} Publishing},
  	volume = {38},
  	number = {15},
  	pages = {153001},
  	author = {Eleonora Di~Valentino and Olga Mena and Supriya Pan and Luca Visinelli and Weiqiang Yang and Alessandro Melchiorri and David F Mota and Adam G Riess and Joseph Silk},
  	title = {In the realm of the Hubble tension{\textemdash}a review of solutions
								            $\less$sup$\greater${\ast}$\less$/sup$\greater$},
  	journal = {Classical and Quantum Gravity}
}

@article{BouPer2024,
  title = {Hubble tension tomography: BAO vs SN Ia distance tension},
  author = {Bousis, Dimitrios and Perivolaropoulos, Leandros},
  journal = {Phys. Rev. D},
  volume = {110},
  issue = {10},
  pages = {103546},
  numpages = {15},
  year = {2024},
  month = {Nov},
  publisher = {American Physical Society},
  doi = {10.1103/PhysRevD.110.103546},
  url = {https://link.aps.org/doi/10.1103/PhysRevD.110.103546}
}

@article{Loubser2025,
    author = {Loubser, S Ilani},
    title = {Measuring the expansion history of the Universe with DESI cosmic chronometers},
    journal = {Monthly Notices of the Royal Astronomical Society},
    volume = {544},
    number = {4},
    pages = {3064-3075},
    year = {2025},
    month = {12},
    issn = {0035-8711},
    doi = {10.1093/mnras/staf1939},
    url = {https://doi.org/10.1093/mnras/staf1939},
    eprint = {https://academic.oup.com/mnras/article-pdf/544/4/3064/65236869/staf1939.pdf},
}

\end{document}